# Bipolaron Model of the Superconductivity of an Iron-Based Layered Compound $LnO_{1-x}F_xFePn$ (Ln =La, Sm, Nd, Pr, Ce, Pn=P, As)

Liang-You Zheng, Bo-Cheng Wang[a], Shan T. Lai[b,c]

Center for Molecular Dynamics and Energy Transfer
Department of Chemistry
The Catholic University of America, Washington, DC 20064, USA

a) Department of Chemistry, Tamkang University, 151 Ying-chuan Road Tamsui, Taipei County Taiwan 25137, ROC
b) Vitreous State Laboratory, the Catholic University of America, Washington, DC 20064, USA
c) To whom requests for reprints should be addressed

*Abstact:*
A bipolaron model is proposed as a superconductivity mechanism for iron-based superconductivity. The results are consistent with the experiments.

## 1. Introduction

The origin of superconductivity in iron-based materials can be studied using a basic theoretical model[1a] such as the bipolaron model, which is introduced in this article. In previous work [1b] it was concluded that one dimensional materials are required to raise the Tc of superconducting materials. Researchers have sought an example of a high Tc iron-based superconductor[2] such as $SmO_{1-x}F_x FeAs$. The Fe-As layer of the crystal structure is shown in Figure 1. A two-site small bipolaron model is proposed as a mechanism for iron-based superconductivity.

The ABCDE and A'B'C'D'E' line segments of the FeAs layer are taken as a quasi-one dimensional chain superconducting transportation pathway as shown in Figure 1.

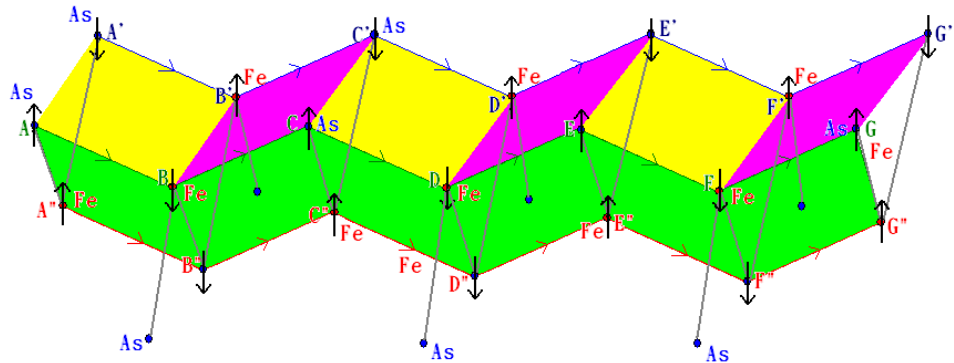

Figure1. Quasi-One Dimensional Model in the Fe-As layer of LnO$_{1-x}$F$_x$FePn (Ln =La, Sm, Nd, Pr or Ce. Pn = P, As) (0<x <1)

## 2. Model Hamiltonian

The object is a two-site small bipolaron in a linear chain, as shown in Fig. 2.

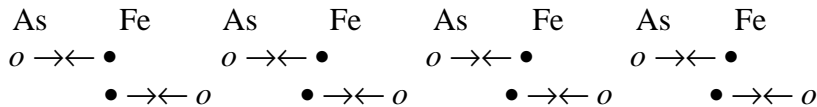

Figure 2. Vibration between Iron and Arsenic

In general, the electron and ion Hamiltonians, $\hat{H}_e$ and $\hat{H}_a$

, respectively, of the system may be written as

$$\hat{H}=\hat{H}_e+\hat{H}_a \qquad (1)$$

where

$$\hat{H}_e=\sum_i h_i=\sum_i\left[\frac{p_i^2}{2m}+\sum_n V(r_i-R_n)\right] \qquad (2)$$

and

$$\hat{H}_a=\sum_n\left[\frac{K}{2}(u_{n+1}-u_n)^2+\frac{M_n}{2}\dot{u}_n^2\right] \qquad (3)$$

In the tight-binding approximation, the electron Hamiltonian reads as

$$\hat{H}_e=-\sum_n t(R_{n+1}-R_n)(C_{n+1}^+C_n+C_n^+C_{n+1}) \qquad (4)$$

where, $t(R_{n+1}-R_n)$ is the interaction of the two nearest neighboring ions with the electron. $R_{n+1}$ and $R_n$ are the instantaneous position of ions. $R_n$ is very small, so as

$$R_{n+1}-R_n=R_{n+1}^{(0)}-R_n^{(0)}+(u_{n+1}-u_n)$$

here

$$u_n=\sum_q\sqrt{\frac{\hbar}{2N M_n\omega_q}}(a_q+a_{-q}^+)e^{i2\pi qna}$$

where, $u_n, u_{n+1}$, are the distances of move of ions from its equilibrium. $a_q^+$ and $a_q$ are the phonon creation and annihilation operators. And the distance of equilibrium position $R_{n+1}^{(0)}-R_n^{(0)}=a$, the lattice constant, is very small, i.e. $u_{n+1}-u_n\langle\langle a$. (The meaning of the previous sentence is unclear, we need to talk about what you mean to state.)

Therefore, the interaction is expanded as

$$t(R_{n+1}-R_n)=t_0-\gamma(u_{n+1}-u_n) \qquad (5)$$

where $t_0 = t(R_{n+1}^{(0)} - R_n^{(0)})$ is the interaction between the electron and the nearest neighboring ions in their equilibrium positions and $\gamma = -\frac{dt}{dx}$ is the rate of the change of the interaction with respect to the distance between ions within the unit cell.

Substituting Eq.(5) into Eq.(4) and taking into account the spin of the electrons, we have

$$\hat{H}_e = -\sum_{n,s}[t_0 - \gamma(u_{n+1} - u_n)](C_{n+1,s}^+ C_{n,s} + C_{n,s}^+ C_{n+1}) \tag{6}$$

The total Hamiltonian is now written as the sum of Eq.(3) and Eq.(6)

$$\hat{H} = \hat{H}_e + \hat{H}_a = \sum_n \frac{K}{2}(u_{n+1} - u_n)^2 + \frac{M_n}{2}\dot{u}_n^2$$

$$-\sum_{n,s}[t_0 - \gamma(u_{n+1,} - u_{n,})](C_{n+1,s}^+ C_{n,s} + C_{n,s}^+ C_{n,s}) \tag{7}$$

Su et al.[3] considered Eq.(5) as the standard form of the electron-phonon coupling in a metal and first applied Eq.(7) to the trans-polyacetylene.

Because we take the two-site small bipolaron models as pair of electrons, one electron spin is up and the other must be down the total spin of the system is zero, i.e. a singlet. Thus, the spin label "s" in the formula is eliminated.

The Hamiltonian is now written as

$$\hat{H} = -\sum_n [t_0 - \gamma(u_{n+1} - u_n)](C_{n+1}^+ C_n + C_n^+ C_{n+1}) + \frac{1}{2}\sum_q \hbar \omega_q \left(a_q^+ a_q + \frac{1}{2}\right) \tag{8}$$

Evaluating the spectrum of the electrons is equivalent to diagonalizing the Hamiltonian operator. To do this the $C_n^+$ and $C_n$ operators are transformed from the Bloch representation to the Wannier representation.

$$C_{n0} = \frac{1}{\sqrt{N}}\sum_k (C_k^0 + C_k^e)e^{-i2\pi kna}$$

$$C_{ne} = \frac{1}{\sqrt{N}}\sum_k (C_k^0 + C_k^e)e^{-i2\pi kna} \tag{9}$$

The Hamiltonian in this representation is given as Equation (10).

$$\hat{H}_e = \sum_k E_0(k)\left(C_k^{0+}C_k^0 + C_k^{e+}C_k^e\right)$$
$$-\sum_k\sum_q B_{k,q}\left(a_q + a_q^+\right)\left(C_k^{0+}C_k^0 + C_k^{e+}C_k^e\right) \tag{10}$$

Where,
$$E_0(k) = -2t_0\cos(2\pi ka) \tag{11}$$

$$B_{k,q} = 2\gamma\cos(2\pi ka)\sqrt{\frac{\hbar}{2N M_n \omega_q}}(i2\pi qa) \tag{12}$$

Eqs. (10) - (12) are valid under the long wave length approximation. In the following, Equations (13) and (14) stem from the fact that the probability of the creation and the annihilation of one electron around odd-number atoms are equal to the probability of the creation and annihilation of one electron around even-number atoms anywhere in the chain (this is true for the two-site small bipolaron).

$$C_k^{e+}C_k^e = C_k^{0+}C_k^0$$
or
$$C_k^{e+}C_k^e + C_k^{0+}C_k^0 = N_k^e + N_k^0 = N_k^B = B_k^+B_k \tag{13}$$

Therefore,
$$\hat{H}_B = \sum_k\left[E_k(0) - H_k\right]B_k^+B_k \tag{14}$$
where
$$\hat{H}_k = \sum_q B_{k,q}\left(a_q + a_{-q}^+\right)$$

The total Hamiltonian of system is rewritten as

$$\hat{H} = \sum_k\left[E_0(k) - H_k\right]B_k^+B_k + \sum_q \hbar\omega_q\left(a_q^+a_q + \frac{1}{2}\right) \tag{15}$$

This is transformed[4] to give equation (16)
$$\hat{H}_T = e^{-S}\hat{H}e^{+S} \tag{16}$$
where

$$S = \sum_k\sum_q \frac{B_{k,q}}{\hbar\omega_q}\left(a_q + a_{-q}^+\right)B_k^+B_k \tag{17}$$

to obtain the Hamiltonian given in equation (18)

$$\hat{H}_T = \sum_k [E_0(k) - \Delta_k] B_k^+ B_k + \sum_q \hbar \omega_q \left( a_q^+ a_q + \frac{1}{2} \right) \qquad (18)$$

where

$$\Delta_k = \sum_q \frac{|B_{k,q}|^2}{\hbar \omega_q} \qquad (19)$$

In equation(18) $\Delta_k$ is the binding energy of two polarons (polaron pair) and also represents the superconductivity gap.

3. Calculation of Tc

Equation (20) is from previous work[5],

$$T_c = \frac{\mu}{k_B} \qquad (20)$$

where $\mu$ is the chemical potential, and $k_B$ is the Boltzmann constant.

Substituting the 3-D value of $\mu$ gives equation (21)

$$k_B T_c = 3.31 \frac{\hbar^2}{m^{**}} n^{2/3} \qquad (21)$$

where $n$ is the concentration of bipolaron, $m^{**}$ is the mass of bipolaron, and $\hbar$ is Planck's constant.

For 2-D situations

$$k_B T_c = \pi \frac{\hbar^2}{m^{**}} n_S \qquad (22)$$

Here $n_S$ is the surface density for two dimensions.

4. Discussion

(1) From measuring the value of $\mu, n, n_S, m^{**}$, it is possible to calculate Tc as shown in the following example:
If n=2E+20 (Low density), $m^{**} = 24 m_e$, then Tc = 42K

(2)     From Figure 1 it is seen  that there are antiferromagnetic chains, ABCDEFG, A'B'C'D'E'F'G', A"B"C"D"E"F"G", and antiferromagnetic blocks, AA'BB', BB'CC', CC'DD', DD'EE', EE'FF', FF'GG', and an antiferromagnetic stripe, AA"BB"CC"DD"EE"FF"GG". Experimentally superconductivity always accompanies antiferromagnetism. These observations are consistent with the experimental trends.

(3)     The properties of $LnO_{1-x}F_xFePn$, are successfully modeled by the quasi one- dimension approach.

References

1)  a) Superconductivity: back to basics
    http://www.rikenresearch.riken.jp/research/594/
    b) L.Y. Zheng, Y.N. Chiu, S. Lai, **THEOCHEM** 722(2005) 147-149.

2)  a)  X.H.Chen, T. Wu, G. Wu,  R.H.Liu, H.Chen, and D.F.Fang,  **Nature**  453( 5 June, 2008 )761-762.
    b) Cao Wang, Linjun Li, Shun Chi, et al. "arXiv: 0804.4290 v2 [Cond-mat-supr-con] 12 June, 2008".
    c) Satoru Matsuishi,Yasunori Inoue, Takatoshi Nomura  et al., **J. Am. Chem. Soc.**  2008, 130(44) 14428-14429.

3)  a)   W. P.Su,  J. Schrieffer, A. Heeger,  **Phys. Rev. Lett.** 42(1979)1698.
    b)   W.P. Su ,J. Schrieffer , A.Heeger, **Phys.Rev.** 22(1980)2099.

4)  G.D. Mahan, **Many Particle Physics**, Plenum , New York  1990.

5)  L.Y. Zheng, Y.N. Chiu, S. Lai, **THEOCHEM**   680(2004)37-39.